\documentclass[twocolumn,prl,showpacs,preprintnumbers,amsmath,amssymb]{revtex4}
\usepackage{graphicx,epsf}
\setkeys{Gin}{width=8.6cm,keepaspectratio}
\usepackage{dcolumn}
\usepackage{bm}
 
\begin{document}


\title{Order-Parameter Criticality of the d=3 Random-Field
Ising Antiferromagnet $\rm Fe_{0.85}Zn_{0.15}F_2$}

\author{F. Ye$^1$,  L. Zhou$^2$, S. Larochelle$^3$, L. Lu$^4$,
D. P. Belanger$^1$, M. Greven$^{4,5}$, and D. Lederman$^6$}

\affiliation{$^1$Department of Physics, University of California,
 Santa Cruz, California 95064}
\affiliation{$^2$T.H. Geballe Laboratory for Advanced Materials, Stanford
 University, Stanford, California 94305}
\affiliation{$^3$Department of Physics, Stanford University, Stanford, 
 California 94305}
\affiliation{$^4$Department of Applied Physics, Stanford University, Stanford,
 California 94305}
\affiliation{$^5$Stanford Synchrotron Radiation Laboratory, Stanford, 
 California 94309}
\affiliation{$^6$Department of Physics, West Virginia University,
 Morgantown, West Virginia 26506}

\date{\today}

\begin{abstract}
The critical exponent $\beta =0.16 \pm 0.02$ for the random-field Ising
model order parameter is determined using extinction-free
magnetic x-ray scattering for $\rm Fe_{0.85}Zn_{0.15}F_2$ in magnetic fields
of $10$ and $11$ T. The observed value is consistent
with other experimental random-field critical exponents,
but disagrees sharply with Monte Carlo and exact ground state
calculations on finite-sized systems.
\end{abstract}


\maketitle

The Ising model is perhaps the most important model in
statistical physics, with many applications that go well beyond the realm of
physics.  In 1944, the pure two-dimensional ($d=2$) model
was solved exactly by Onsager \cite{o44}.  For the pure $d=3$ Ising model,
although various calculations and computer simulation techniques have
proven extremely useful, no exact results are available.
Nevertheless, there is extremely good agreement among theory,
simulations, and experiments \cite{b00}.  Hence,
the model can be considered well understood.
Through the years, the intellectual value of the Ising model
has grown, particularly as a model of disorder.  One of the most important of
these models of disorder occurs when a random field is imposed which
couples directly to the order parameter of the system. The most studied
realization of this random-field Ising model (RFIM) is the diluted
antiferromagnet in an applied magnetic field. Unlike the pure $d=3$
Ising model, there is, so far, poor agreement between theory and
simulations on the one hand and experiments on the other \cite{b00}.
This has motivated us to measure the critical behavior of the
staggered magnetization, the antiferromagnetic order parameter
\begin{equation}
M_s=M_0t^{\beta}
\end{equation}
where $t=(T_c(H)-T)/T_c(H)$, of the dilute Ising
antiferromagnet in an applied field, since this is one of the most 
valuable yet also 
least characterized aspects of the experimental system.
Our result for the order parameter exponent $\beta$ provides an important
quantitative experimental contribution toward a comprehensive understanding
of the RFIM.

The dilute, insulating antiferromagnet
$\rm Fe_xZn_{1-x}F_2$ has a large
single-ion anisotropy and is an extensively studied \cite{b00}
$d=3$ RFIM realization \cite{fa79,c84}.
Nevertheless, prior attempts to determine the critical behavior
of the order parameter have been unsuccessful.  This
may be surprising since, in principle, one only
needs to measure the temperature dependence of the
Bragg scattering intensity, $I_B$, which is
proportional to $(M_s)^2$, with a magnetic field $H$
applied along the $c$-axis, the spin ordering direction.
For two reasons, such measurements have proven
very difficult in practice.
First, neutron scattering on these high-quality
crystals suffers greatly from the effects of extinction;
the beam, upon transmission through the crystal,
is depleted of neutrons satisfying
the Bragg condition, resulting in the saturation
of the measured value of $I_B$. 
Second, for magnetic concentrations $x$
below the vacancy percolation threshold \cite{bb00,samb02},
$x_v = 0.754$,
domain formation obscures
the RFIM critical behavior below the transition
at $T_c(H)$ in $\rm Fe_xZn_{1-x}F_2$
and its less anisotropic isomorph
$\rm Mn_xZn_{1-x}F_2$.  Although the domains
may be internally well ordered,
$I_B$ will be greatly diminished
if the characteristic length
scale for the domain structure is smaller than that of
the spectrometer resolution; we will refer to this as
micro-domain structure and it has been studied extensively
in previous works \cite{n93}.
Under severe extinction conditions,
domain structure may relieve
extinction and actually cause $I_B$
to increase. Whether domain structure forms or not,
the Bragg scattering cross section will be decreased
by thermal disorder as the transition
is approached.  In x-ray scattering, since the magnetic scattering
cross section is relatively small, the scattering intensity, obtained
in a reflection geometry, does not suffer from extinction,
as extensively discussed previously \cite{gmshgptb87,hfbt93}.
The use of extinction-free magnetic x-ray scattering, and 
of a crystal with $x>x_v$ to avoid
micro-domains, has allowed us to accurately characterize
the order parameter critical behavior in $\rm Fe_{0.85}Zn_{0.15}F_2$.

The magnetic x-ray scattering technique was employed for
$\rm MnF_2$ for $H=0$
by Goldman {\it et al.\ }\cite{gmshgptb87}, and was then
applied to $\rm Mn_xZn_{1-x}F_2$ with $H>0$
by Hill {\it et al.\ }\cite{hfbt93}.  Whereas the $H=0$ study 
yielded the exponent $\beta$
consistent with the
$d=3$ Ising model, the latter did not
reveal the universal RFIM behavior \cite{hfbt93},
which would be consistent with $x<x_v$;
the $H=0$ transition temperature \cite{bkbj80}
and its field dependence \cite{rkj88} are consistent with the
concentration being very close to or below $x_v$.  That the transition
in Ref. \cite{hfbt93} was obscured by micro-domains \cite{bb00}
is suggested by the zero slope of
${M_s}^2$ versus $T$ as $T \rightarrow T_c(H)$.  The present
magnetic x-ray scattering measurements use $\rm Fe_{0.85}Zn_{0.15}F_2$,
for which $x$ is well above $x_v$.

The measurements were made at the new high-field magnet facility
on beam line 7-2 of the Stanford Synchrotron Radiation Laboratory.
A monochromatic x-ray beam was obtained from the wiggler spectrum
via a Si(111) double-crystal monochromator. X-ray energies between
14 and 13.5 keV were used,
which resulted in a penetration depth of about 60 $\mu$m.
The energy was tuned to minimize energy-sensitive
multiple scattering \cite{hfbt93}.
The sample had a finely polished face, a few
mm$^2$ in area, with the $a$-axis perpendicular
to the polished face and the $c$-axis along the vertical
field.  The temperature of the crystal, mounted in
a He atmosphere, was stable to approximately 10 mK.
The transition temperature for $H=0$ was measured to
be $T_N=66.7$~K, consistent with birefringence measurements
on the same sample \cite{yb02} and with a concentration
$x=0.85$ \cite{bkbj80}. 
It has been shown that the antiferromagnetic
transition at this magnetic concentration is stable
at fields as high as $H=18$~T \cite{samb02}.
For $H=10$ and $11$~T, the transitions
are at $T_c=64.2$ and $64.0$~K, respectively.  The
lattice constants of the sample were determined to be
approximately $a=4.68$ \AA\ and $c=3.27$ \AA\ near the transition
temperature.  The half-widths-at-half-maximum for the
Bragg peaks were $4 \times 10^{-4}$, $4 \times 10^{-3}$,
and $4 \times 10^{-3}$ reciprocal lattice units
for the transverse, longitudinal and vertical
directions, respectively, at the (100)
magnetic Bragg point, about which transverse $H=0$
and $H=11$~T and longitudinal $H=10$~T scans were obtained. 
The sample was remounted between measurements at
different fields, and therefore we normalized intensities using
scans at $T=47$~K.
Three conventional thermal-cycling procedures were
employed.  In ZFC, the sample is cooled in zero field
below $T_c(H)$, the field is
applied, and the sample is warmed through
$T_c(H)$, waiting at each temperature at least
20 min before taking data to let the
temperature and system stabilize.
In FC, the sample is cooled through
$T_c(H)$ in the field, taking data as in ZFC. 
Field-heated (FH) data were taken by heating in the field after FC.
The scans typically consisted 
of 41 points, about 15 of which covered the Bragg peak.
At each point, the intensity was counted for 30 to 45 seconds,
depending on the temperature of the scan.

\begin{figure}
 \centerline{
  \epsfxsize=9cm 
   \epsfbox{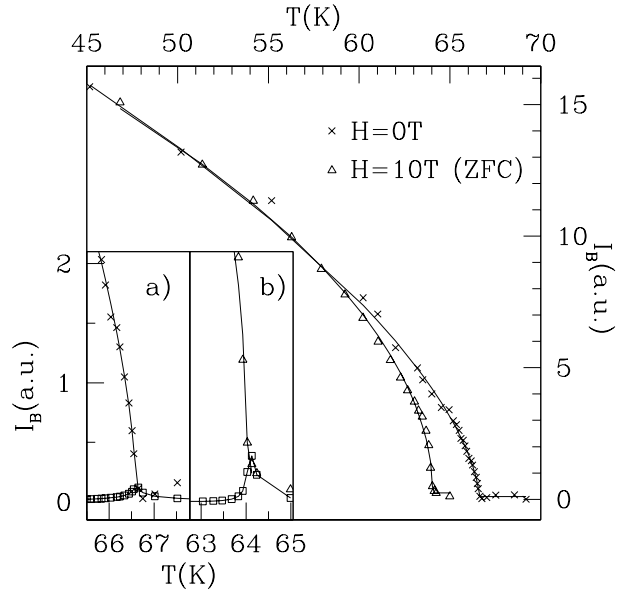}
 }
\caption{The ZFC Bragg intensity, $I_B$, in arbitrary units (a.u.),
versus $T$ for $H=0$ and $H=10$~T,
with the $T$-independent background
intensity subtracted.  The square symbols in insets a) and b)
show the critical scattering contributions to the x-ray intensities
for  $H=0$ and $H=10$~T, respectively,
determined from neutron scattering measurements as described
in the text.}
\end{figure}

Figure 1 shows the Bragg intensity for $H=0$ and
$10$~T versus temperature, with the $q$ and $T$ independent
backgrounds subtracted, where $q$ is the distance
in reciprocal space from the (100) antiferromagnetic Bragg
point.  The background depends on the
precise experimental configuration, but not on the
thermal cycling used to collect data, and is mostly from
sources other than the crystal itself.  For comparison of the background
to the Bragg signal, typical background
counts for the $H=11$~T scans were
eight counts per second whereas the $q=0$
intensity was 160 counts per second at $T=47$~K.
Above the transition, the scattering intensity results
only from the critical scattering and goes
to zero well above $T_c(H)$, indicating that there are
no contributions from multiple scattering to the
Bragg intensity.  To determine the
critical scattering for $H=0$, $10$ and $11$ T,
neutron scattering line shapes, obtained with a sample of
nearly the same magnetic concentration \cite{ymkyfb02} using
a previously described procedure \cite{sbf99}, were
folded with the x-ray resolution, and the overall $q=0$ amplitude
was adjusted to fit the $H=10$~T data above $T_c(H)$.
Insets a) and b) in Fig.\ 1 show the critical scattering
contributions for $H=0$ and $10$~T, respectively.
As a result of the high momentum
resolution of the x-ray technique,
the critical scattering contributions, which are
nearly Lorentzian
for $H=0$ \cite{sbf99,bybf02}, are almost negligible
(Fig. 1a)).  For $H>0$, however, the critical scattering
has a much larger $q$ dependence \cite{sbf99} at small $q$.
Consequently, a small contribution to the $q=0$ scattering
is more discernible for the $H=10$~T data.
These contributions were subtracted from all the data
before determining the order parameter exponent.

Although neutron scattering measurements
using $\rm Fe_{0.85}Zn_{0.15}F_2$ \cite{ymkyfb02}
and $\rm Fe_{0.93}Zn_{0.07}F_2$ \cite{sbf99} show no evidence
for micro-domain formation in
the  critical scattering, $H>0$ hysteresis in $I_B$ is evident, with the
FC intensities larger than the ZFC ones, a result
of extinction.  The x-ray Bragg scattering also
shows hysteresis, but in this extinction-free case
the ZFC data are higher in intensity.  
In the random field region, the FC intensity is factor of
4.1 smaller than ZFC intensity. This ratio depends slightly
on the cooling rate used in obtaining the FC data. The ZFC
data are rate independent. For $H=11$~T,
the corresponding
ratio is approximately $4.0$.  FH data were intermediate between
the ZFC and FC curves.   We note that specific heat measurements
also show hysteresis
very close to $T_c(H)$ at this concentration \cite{yb02,sb98}.  Such
hysteresis is likely a result of the extremely slow activated
RFIM dynamics and possibly represents the fact that FC long-range
ordering must take place while traversing the transition, where
activated dynamics plays the greatest role.
The logarithmically slow relaxation associated with
activated dynamics \cite{f86,kmj86} severely limits the ability
of the system to equilibrate extremely close to $T_c(H)$.
This limits the formation of long-range order upon FC.  ZFC data,
on the other hand, are obtained without approaching $T_c(H)$
except when the order parameter is already very small, and
thus do not visibly suffer from the slow dynamics.  Moreover,
no dependence on the rate of temperature change was observed
upon ZFC.  Hence, we believe the ZFC data represent the correct order
parameter measurement.  
Various measurements near the transition at this concentration
have yielded critical behavior indicative of a second-order phase
transition.  There is also no measureable latent-heat in specific
heat critical behavior measurements \cite{yb02}. 
Therefore, this appears to be a
second-order transition, although an extraordinary one.

The normalized Bragg intensity curves in Fig.\ 1 clearly approach
$T_c(H)$ vertically.  This is characteristic
of experiments \cite{ymkyfb02,sbf99,bybf02} and
simulations \cite{bb00} for $x>x_v$ and in stark contrast with
experiments \cite{b00,hfbt93,bwshnlrl96} and simulations \cite{bb00}
for $x<x_v$,
where $I_B$ approaches $T_c(H)$ horizontally.  The latter behavior
is attributable to micro-domain formation, which is
energetically favorable when the vacancies percolate through
the crystal, as shown in Monte Carlo simulations \cite{bb00}.

\begin{figure}
 \centerline{
  \epsfxsize=8.5cm 
   \epsfbox{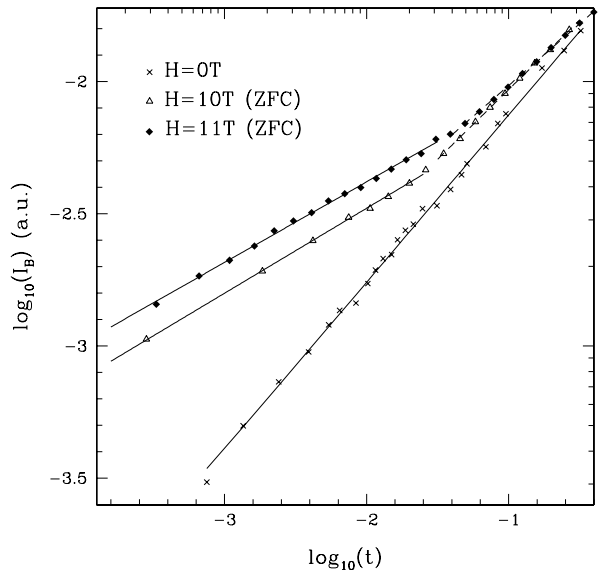}
 \vspace*{3mm}
 }
\caption{The same ZFC data as in Fig.\ 1 as well as data taken
at $H=11$~T, corrected for
the critical scattering contribution, plotted as
the logarithm of the intensity versus the logarithm of $t$.
The solid lines for $H=10$ and 11 T indicate RFIM behavior
with $\beta = 0.16$, while the solid line for $H=0$
reflects conventional random-exchange behavior ($\beta = 0.35$).
}
\end{figure}

Figure 2 shows the logarithm of $I_B$, with the constant
background and critical scattering contributions subtracted,
for $H=0$, $10$, and $11$ T, versus the logarithm of $t$.
The values of $T_c(H)$ were determined from fits to the data.
For $0.0007<t<0.03$ and $H=0$, we find $\beta = 0.35 \pm 0.02$
(solid line),
which agrees well with
several experimental and theoretical determinations for the random-exchange
Ising model \cite{b00}.
For $H=10$ and $11$~T,
a crossover from random-exchange to RFIM critical behavior occurs
near $t=0.03$, consistent with
birefringence measurements \cite{yb02},
and the data can be fit to a single power law only
in the range $0.0001<t<0.03$.  The fits over this range
yield the exponent $\beta = 0.16 \pm 0.02$
for the combined $H=10$ and $11$~T data and are
indicated by the parallel solid lines in Fig. 2.
A less sophisticated data analysis, which did not
correct for the critical scattering contribution, resulted in 
a value for the order parameter exponent that is larger by 0.02, still
within the error bars.  The correction should, of course, be done 
in order to obtain the correct value of $\beta$.  The slope of the data at
large $t$ changes with $H$ since the definition of the 
reduced temperature involves 
$T_c(H)$ for $H>0$ and not the $H=0$ transition temperature $T_N$.

Through the Rushbrooke scaling relation
\begin{equation}
2\beta + \gamma + \alpha \ge 2 \quad ,
\end{equation}
which is usually satisfied as an equality,
$\beta$ is related to the universal critical
exponents $\alpha$ (for the specific heat) and
$\gamma$ (for the staggered
susceptibility) of the $d=3$ RFIM.
The experimentally determined specific heat peak is nearly logarithmic
and very symmetric close to $T_c(H)$,
consistent with $\alpha \approx 0$ \cite{yb02,sb98}.
Neutron scattering analyses \cite{ymkyfb02,sbf99}
yield values in the $1.45 < \gamma < 1.65$ range.  Therefore, 
the experimental
value $\beta \approx 0.16$ is fairly consistent with Rushbrooke scaling,
taking the upper limit of $\gamma$ and $\alpha =0$.

A very recent NMR study \cite{kdlbzp02}
of the order parameter in the effective
short-range interaction random-field
ferroelectric $\rm Sr_{0.61-x}Ce_xBa_{0.39}Nb_2O_6$, with
$x=0.0066$, yielded $\beta = 0.14 \pm 0.03$,
consistent with the present result for
$\rm Fe_{0.85}Zn_{0.15}F_2$.  It was obtained, however, in
the presence of micro-domain structure.  Apparently,
the more local NMR probe is less sensitive
than the Bragg scattering techniques to the formation
of micro-domains.  This suggests that
domain formation does not preclude a fairly
sharp RFIM-like phase transition and a
measurement of its order parameter \cite {rkjr88},
but only prevents measurements of
the order parameter through scattering experiments
for $x<x_v$.

There does not exist a set of theoretical
results that are consistent with all the
experiments \cite{b00}.  Monte Carlo \cite{r95} and exact
ground state calculations \cite{hy01} yield very
small values for $\beta$ and large, negative
values for $\alpha$.  Other numerical and scaling analyses \cite{mf01}
yield $\alpha$ close to zero, consistent with
experiments \cite{yb02,sb98}, but also yield $\nu=1.37 \pm 0.09$,
much larger than experimental value
$\nu=1.05 \pm 0.01$ \cite{sbf99,ymkyfb02}.  Another
recent work \cite{dm02} yields $\alpha$ and $\beta$ close to zero.
One Monte Carlo study, on a large lattice and with
less assurance of equilibrium than other simulations,
yielded $\beta = 0.25 \pm 0.03$ \cite{bb00b}.
Since consistency among numerical and
experimental exponents continues to elude us,
a comprehensive understanding of the $d=3$ RFIM is yet
to be achieved. The determination of the order parameter
exponent presented here is an important quantitative 
contribution in this direction.

We thank M. Matsuda, S. Katano, H. Yoshizawa
and J. A. Fernandez-Baca for allowing the use of unpublished neutron scattering
results in the analysis shown in Fig.\ 1, and the SSRL staff for their
help in building the magnet facility.
The x-ray experiments were carried out at the Stanford Synchrotron
Radiation Laboratory, a national user facility operated by Stanford
University on behalf of the U.S. Department of Energy, Office of Basic
Energy Sciences. The work at Stanford was also supported by the U.S.
Department of Energy under Contract Nos. DE-FG03-99ER45773 and
DE-AC03-76SF00515, by NSF Grant Nos. DMR-9985067 and DMR-9802737, and by
the A.P. Sloan Foundation. The work at UCSC was supported by Department of
Energy Grant No. DE-FG03-87ER45324. The work at West Virginia University
was supported by NSF Grant No. DMR-9734051.

\end{document}